# Knowledge Representation Concepts for Automated SLA Management

*Adrian Paschke and Martin Bichler[1]*

Abstract— Outsourcing of complex IT infrastructure to IT service providers has increased substantially during the past years. IT service providers must be able to fulfil their service-quality commitments based upon pre-defined Service Level Agreements (SLAs) with the service customer. They need to manage, execute and maintain thousands of SLAs for different customers and different types of services, which needs new levels of flexibility and automation not available with the current technology. The complexity of contractual logic in SLAs requires new forms of knowledge representation to automatically draw inferences and execute contractual agreements. A logic-based approach provides several advantages including automated rule chaining allowing for compact knowledge representation as well as flexibility to adapt to rapidly changing business requirements. We suggest adequate logical formalisms for representation and enforcement of SLA rules and describe a proof-of-concept implementation. The article describes selected formalisms of the ContractLog KR and their adequacy for automated SLA management and presents results of experiments to demonstrate flexibility and scalability of the approach.

## 1. Introduction

Outsourcing of complex IT infrastructure to IT service providers has become increasingly popular and led to much recent development [1]. To ensure Quality of Service (QoS) between the service customer and the provider, they jointly define a Service Level Agreement (SLA) as a part of a service contract that can be monitored by one, both or third parties. An SLA provides metrics for measuring the performance of the agreed upon Service Level Objective (SLOs). SLA rules are used to monitor service execution and detect violations of SLOs. SLAs are fundamental not only for outsourcing relationships, but for any kind of service supply chain and assume a central position in popular IT service management standards such as ITIL (www.itil.co.uk) or BS15000 (www.bs15000.org.uk). As a consequence, IT service providers need to manage, execute and maintain thousands of SLAs for different customers and different types of services in the upcoming service-oriented computing landscape. Commercial service level management tools such as IBM Tivoli, HP OpenView, CA Unicenter, BMC Patrol, and Microsoft Application Center store se-

---

[1] A. Paschke and M. Bichler are with the Internet-based Information Systems, Department of Informatics, TU München, Boltzmannstr. 3, D-85748 Garching. E-mail: {Paschke|Bichler}@in.tum.de.

Paschke, A. and Bichler, M.: Knowledge Representation Concepts for Automated SLA Management, Int. Journal of Decision Support Systems (DSS), submitted 19th March 2006.

lected QoS parameters such as availability or response time as parameters in the application code or database tiers. This approach is restricted to simple, static rules with only a limited set of parameters. We have analyzed a large number of text-only real-world SLAs [2] and found various types of complex rules which need to be enforced by IT service providers such as graduate rules, dependent rules, normative rules, default rules, exception rules. In most cases these SLA rules are part of a hierarchical set of contracts, consisting of a basic agreement with general terms and conditions, a group-level service agreement, one or more SLAs and internal operation level agreements (OLAs) or underpinning contracts [2]. These types of interlinked, unitized rule sets with multiple conditions which are possibly scattered among business partners, organizations or departments demand for a more sophisticated knowledge representation (KR), which allows an IT service provider to analyze, interchange, manage and enforce large amounts of (decentralized) rules and adapt them during runtime. Moreover, traceability and verifiability of drawn conclusions (e.g. derived penalties) and triggered reactions during contract monitoring and enforcement is a requirement in order to fulfil legal regulations and compliance rules.

In this article we propose a declarative rule-based approach to SLA representation and management. Whereas, existing approaches to standardize SLAs are based on imperative procedural or simple propositional logic (see section 6), we draw on logic programming (LP) and related knowledge representation concepts. LP allows for a compact knowledge representation of SLA rules and for automated rule chaining by resolution and variable unification, which alleviates the burden of having to implement extensive control flows as in imperative programming languages and allows for easy extensibility. However, further logical formalisms are needed for automated SLA management. We propose an expressive framework of adequate knowledge representation (KR) concepts, called ContractLog, which combines a labelled, typed and prioritized logic, reactive rules and complex event processing, temporal event logics (event calculus), defeasible logic, deontic logic, description logic and verification and validation techniques for the domain of SLA management. In particular, the integration of reactive and derivation rules with complex event/action processing, KR based event logics, deontic norms and procedural attachments enables an adequate representation of contractual rules as they can be found in SLAs nowadays. We describe an implementation of the ContractLog framework and show how the different types of KR concepts can be transformed into general LPs and LP based meta programs. ContractLog enables a clean separation of

Paschke, A. and Bichler, M.: Knowledge Representation Concepts for Automated SLA Management, Int. Journal of Decision Support Systems (DSS), submitted 19[th] March 2006.

concerns by specifying contractual logic in a formal, declarative machine-readable and executable fashion and enables calls to imperative object-oriented code (Java) or existing system monitoring tools. In this article, we describe adequate knowledge representation concepts for SLA Management and provide an experimental evaluation of the approach focusing on scalability and flexibility based on real-world use cases. Our research follows the design science research in IS approach as described in Hevner et al. [3]

The article is organised as follows: We first provide an overview of SLAs in section 2, define relevant terms and introduce a use case based on real-world industrial SLAs. In section 3 we describe the ContractLog KR and the selected logical formalisms for the adequate representation of SLAs. In section 4 we introduce superimposed rule-based service level management tools (RBSLM) which serves as a proof-of-concept implementation. Section 5 presents experimental evaluations and illustrates the rule-based formalization of SLA rules in the ContractLog KR based on the use case example defined in section 3. Section 6 discusses related work. Finally, section 7 concludes with a short summary of key findings.

## 2. Service Level Agreements

An SLA is a document that describes the performance criteria a provider promises to meet while delivering a service. It typically sets out the remedial actions and penalties that will take effect if performance falls below the agreed service levels. It is an essential component of the legal contract between a service consumer and the provider. In the following, we will describe relevant terms, before we introduce a use case that we will use later for the evaluation.

### 2.1 Terminology

*SLA rules* represent guarantees with respect to graduated high/low ranges of metrics (e.g., average availability range [low: 95% , high: 99%, median: 97%]) so that it can be seen whether the measured metrics exceed, meet or fall below the defined service levels during a certain time interval. They can be informally represented as rules which might be chained in order to form graduations, complex policies and conditional guarantees, e.g., *"if the average service availability during one month is below 95% then the service provider is obliged to pay a penalty of 20% of the monthly service fee"*.

Paschke, A. and Bichler, M.: Knowledge Representation Concepts for Automated SLA Management, Int. Journal of Decision Support Systems (DSS), submitted 19[th] March 2006.

According to their intended purpose, their scope of application or their versatility SLAs can be grouped into different categories. In this paper, we will use the following terms described in Table 1.

TABLE 1: SLA CATEGORIZATION

| Purpose of the Contract | |
|---|---|
| Basic Agreement | Defines the general framework for the contractual relationship and is the basis for all subsequent SLAs. |
| Group-level Service Agreement | Subsumes all components which apply to several subordinated SLAs. |
| Service Level Agreement | Main contract between service provider and service customer. |
| Operation Level Agreement (OLA) | A contract with internal operational partners, needed to fulfil an SLA. |
| Underpinning Contract (UC) | A contract with an external operational partner, needed to fulfil an SLA. |

Service Level Agreements come in several varieties and comprise different technical, organizational or legal components. Table 2 lists some typical contents.

TABLE 2: CATEGORIZATION OF SLA CONTENTS

| Technical Components | Organizational Components | Legal Components |
|---|---|---|
| - Service description<br>- QoS metrics<br>- Actions<br>- … | - Liability / liability limitations<br>- Level of escalation<br>- Maintenance periods<br>- Monitoring and reporting<br>- Change management<br>- … | - Obligations to co-operate<br>- Legal responsibilities<br>- Proprietary rights<br>- Modes of invoicing and payment<br>- … |

Although the characteristics and clauses may differ considerably among different contracts, they all include more or less static parts such as the involved parties, the contract validity period, the service definitions but also dynamic parts which are more likely to change, such as the QoS definitions stated as SLA rules specifying service level guarantees and appropriated actions to be taken if a contract violation has been detected according to measured performance values via SLA metrics. The representation of the static part of an SLA is straightforward. From the point of view of a rule-based decision/contract logic they can be simply represented as facts managed e.g. in the knowledge base or a external database or Semantic Web ontology (A-Box). In this article we focus on the dynamic part of an SLA – the SLA rules.

**2.2 SLA Use Case**

In order to better illustrate the requirements of SLA management we will briefly describe a use case derived from real-world SLA examples from and industry partner The SLA defines three monitoring schedules, "*Prime*", "*Standard*" and "*Maintenance*":

Paschke, A. and Bichler, M.: Knowledge Representation Concepts for Automated SLA Management, Int. Journal of Decision Support Systems (DSS), submitted 19[th] March 2006.

TABLE 3: MONITORING SCHEDULES

| Schedule | Time | Availability | Response Time |
|---|---|---|---|
| Prime | 8 a.m. -18 p.m. | 98%[99%]100% ; pinged every 10s | 4 sec.; pinged every 10s |
| Standard | 18 p.m. -8 a.m. | 95%[97%]99%; pinged every min. | 10[14]16 sec.; pinged every min. |
| Maintenance | 0 a.m.- 4 a.m.* | 20%[50%]80%;pinged every 10 min | No monitoring |

During prime time the *average availability* has a low value of *98%,* a median of *99%* and a high value of *100%* and a *response time* which must be below *4s*. The service metrics are calculated via a ping every 10 seconds. During standard time the average availability is *{high:99%;low:95%;median:97%}* and response time *{high:10sec.;low:16sec.;median:14sec.}* monitored via a ping every minute. Maintenance is permitted to take place between midnight and 4a.m. During this period the average availability is *{high:80%; low:20%; median:50%}* monitored every 10 minutes. Response time will not be monitored in this case. Further the SLA defines a "bonus-malus" policy:

TABLE 4: "BONUS MALUS" PRICE POLICY

| Price | Base | Bonus | Malus |
|---|---|---|---|
| Prime | $p_{prime}$ | $p_{prime} + (X_{high}-X_{median}) * p_{bonus}$ % | $p_{prime} - (X_{median}-X_{low}) * p_{malus}$ % |
| Standard | $p_{standard}$ | $p_{standard} + (X_{high}-X_{median}) * p_{bonus}$ % | $p_{standard} - (X_{median}-X_{low}) * p_{malus}$ % |
| Maintenance | $p_{maintenance}$ | $p_{maintenance} + (X_{high}-X_{median}) * p_{bonus}$ % | $p_{maintenance} - (X_{median}-X_{low}) * p_{malus}$ % |
| Incident Penalty | - $p_{incident}^{n}$ | | |

According to the monitoring schedules a differentiated base price is defined if the service levels are met. If the service levels are exceeded (median to high) a dependent bonus is added and if they fall below the agreed upon service levels (median to low) a discount is deducted. The bonus and malus are defined as a percentage value of the base price. If a service level is missed, i.e. the actual value falls below the low service level (<low) an additional penalty has to be paid which increases exponentially with the number of incidents during the accounting period. In case of outages/incidents the SLA defines two escalation levels:

TABLE 5: ESCALATION LEVELS WITH ROLE MODELS AND ASSOCIATED RIGHTS AND OBLIGATIONS

| Level | Role | Time-to-Repair (TTR) | Rights / Obligations |
|---|---|---|---|
| 1 | Process Manager | 10 Min. | Start / Stop Service |
| 2 | Quality Manager | Max. Time-to-Repair (MTTR) | Change Service Levels |

Each escalation level defines clear responsibilities in terms of associated roles which have certain rights and are obliged to do certain remedial actions in case of incidents which initiate the respective escalation level. In the SLAs' escalation level 1 the process manager is obliged to restart an unavailable service within 10 minutes. Accordingly, she has the right (permission) to start and stop the service. If she fails to do so, escalation level 2 is triggered and the quality manager is informed. The quality manager has more rights, e.g. the right (permission) to adapt/change the SLA management systems respectively the service levels. The quality manager might discuss the time needed to repair with the process manager and extend it up to a *maximum time to repair* level (change request). In case of very critical incidents the system might directly proceed to escalation level 2 and skip level 1.

## 3. ContractLog

ContractLog [4] is an expressive and computationally feasible KR framework consisting of adequate KR concepts used to describe contracts and SLAs resp. It combines selected logical formalisms which are all implemented on the basis of logic programming (mainly a meta programming approach based on derivation rules). It provides a typed, labelled, unitized and prioritized logic with extended Java-based procedural attachments which enable reuse of external (procedural) functionalities, tools and data directly into declarative LPs execution. In the following, we describe the core syntax and semantics of ContractLog LPs in the context of logic programming. We then elaborate on advanced (non-monotonic) KR concepts, which are required to adequately formalize typical SLA rules.

### 3.1 Notation and Semantics of ContractLog

In this article we use the standard LP notation and extended logic programs with default negation and explicit negation for the knowledge base. We assume that the reader is familiar with basic Horn theory and logic programming. For more information on the major theoretical results and implementations w.r.t. Prolog we refer the reader to e.g. [5] and [6].

**Core Syntax of ContractLog**

A ContractLog LP is an e*xtended LP* (ELP). An ELP is a set of clauses (rules) of the from $H \leftarrow B$, where $H$ is a literal over $L$ called the *head* of the *derivation rule*, and $B$ is a set of literals over $L$ called the *body* of the rule. A literal is either an atom or the negation "~" resp. "¬" of an atom, where "~" is denoted as *default negation* and "¬" as *explicit negation*. Roughly, default negation means, everything that can not be proven as true is assumed to be false. Its implementation is given in the usual way by a negation-as-finite-failure rule. A rule is called a *fact* if it only consists of the rule head $H \leftarrow$. An atom is a n-ary formula

Paschke, A. and Bichler, M.: Knowledge Representation Concepts for Automated SLA Management, Int. Journal of Decision Support Systems (DSS), submitted 19[th] March 2006.

containing terms *p(a, X, f(...))*, where *p* is the predicate name. A term is either a constant *a*, a (possibly free) variable *X* or a n-ary complex term/function *f(...)*. A goal/query *G?* is a headless clause defining a conjunction of literals (positive or negative atoms) ← $L_1$ ∧ .. ∧ $L_i$ where each $L_i$ is called a subgoal. The ContractLog KR uses an extended ISO Prolog related scripting syntax (ISO Prolog ISO/IEC 13211-1:1995) called Prova (http://www.prova.ws/ [7]) to write ContractLog LPs as scripts, where a variable starts with a upper-case letter, e.g. *X,Y,Z*, a constant/individual with a lower-case letter, e.g. *a,b,c* and a query is written as a function *:-sovle(...)* or *:-eval(...)*, ← is denoted by ":-" and ∧ by ";". Default negation is written as *not(...)*, e.g. *not(p())*, and explicit negation as *neg(...)*, e.g. *neg(p())*.

**Configurable and Selectable Semantics of ContractLog**

The inference engine coming with ContractLog is a *configurable* engine with different selectable semantics and test suites to verify and validate correctness of execution of arbitrary (possibly distributed and interchanged) LPs. The ContractLog/Prova engine has been developed based on the Mandarax derivation rule engine [8] and the Prova inference and language extensions [7] and has been integrated into the Prova 2.0 distribution [8]. It supports different inference features and several semantics which can be selected and configured. The particular advantage of this approach is, that according to the respective logic class of a ContractLog LP which should be executed, e.g. definite LP without negation or generalized LPs with negation-as-failure (but without explicit negation), a less expressive and hence more efficient semantics in terms of performance for query answering might be chosen and properties such as memory consumption w.r.t. goal memoization or weakening vs. safeguarding of decidability w.r.t. safety conditions (e.g., Datalog) can be configured. For example, the basic 2-valued SLDNF resolution with the negation-as-finite-failure rule which is the weakest "semantics" supported by ContractLog does not support memoization and suffers from well-known problems such as endless loops or floundering. It is not complete for LPs with negation or infinite functions. Moreover, it can not answer free variables in negative subgoals since the negation as finite failure rules is only a simple test. For more information on SLDNF-resolution we refer to [9, 10]. For typical unsolvable problems related to SLDNF see e.g. [11]. In contrast, (extended) 3-valued well-founded semantics (WFS) which is the strongest semantics supported by ContractLog is fully declarative, decidable and has polynomial worst-case complexity for several logic classes. The *well-founded semantics* (WFS) of Van Gelder et. al. [12] is a 3-valued semantics with three possible truth val-ues: *true* (t), *false* (f) and *unkown* (u). We refer to [12-15] for a definition. The procedural semantics, called SLE resolution (**L**inear resolution with **S**election function for **E**xtended WFS), implemented in ContractLog to compute WFS extends linear SLDNF with goal memoization and loop prevention. The major difference to tabling-based approaches such as SLG-resolution [16, 17] is that it preserves the linearity property of the resolution algorithm with sequential tree-based formulations like in SLDNF and hence still enables efficient stack-based memory structures and expressive sequential operators.

**Representation of Contract Rules / Business Rules**

Before we elaborate on further (non-monotonic) logical formalisms and rule types which are needed for adequately formalising SLAs, we will briefly demonstrate the general applicability of derivation rules (as described above) for the representation of contract rules and in particular for the representation of SLA rules and higher-level policy rules. SLA rules typically have the form "*if ... then ...(else)*". Such informal rules can be formalized as a set of prerequisites (conditions) which form the body of a derivation rule and a conclusion (consequent) which forms the head of the rule. The rules are relatively easy to write since the user only needs to express *what* (decision logic) they want. The responsibility to interpret this and to decide on *how* to do it is delegated to an interpreter (an inference engine). Table 6 gives an example of the translation of an informal business rule set, which might occur in a SLA, to a formal representation using the Prolog related syntax introduced previously.

TABLE 6: EXAMPLE OF A BUSINESS RULE SET

| |
|---|
| (r1) "If customer has spend more than 1000$ in the last year then customer is a bronze customer. (r2) "If the customer is a bronze customer he will get a discount of 5%." |
| (r1) discount(Customer, 5%) :- bronze(Customer). (r2) bronze(Customer) :- spending(Customer, Value, last year) , Value >1000. |
| Fact:    spending(Peter Miller, 1200, last year). |

Negation can be formalized with either default or explicit negation, depending on the intention. For example, rules with epistemic character such as "*if X is not believed to be a valuable customer, then X is assigned standard level*" might be formalized with default negation as follows: *standard(Customer) :- not(spending(Customer, >1000$, last year)*, i.e. the rule expresses a default rule which holds in case there is no information about the spending or the spending of the customer is below 1000$ in the last year. The else part of an informal rule can be represented in ContractLog either using the sequential operator cut "!", e.g. "*h(X) :- b(X), !.   h(X):-c(X).*" or using negation-as-failure, e.g. "*h(X) :- b(X). h(X):-not(b(X)), c(X).*".

Paschke, A. and Bichler, M.: Knowledge Representation Concepts for Automated SLA Management, Int. Journal of Decision Support Systems (DSS), submitted 19[th] March 2006.

The former approach in many cases results in efficient computations, but imposes a procedural character with partially ordered rules on the LP. The later approach is more declarative, but might need redundant goal computations if the used semantics does not support goal memoization. In general, derivation rules can be executed using forward reasoning or using goal-driven backward-reasoning. There exist efficient forward-reasoning production rule engines (Condition→Action rules) such as Jess or Drools which are based on the Rete algorithm. However, these systems typically have a rather restricted expressiveness and only an operational semantics, but no clear logical semantics which is crucial in the context of verifiable, traceable and highly reliable derived results which should count even in the legal sense of an SLA. Moreover, due to the used algorithms in forward chaining systems lots of rules would be eligible to fire in any cycle and a lot of irrelevant conclusions are drawn. In backward-reasoning the knowledge base can be temporarily populated with the needed extensional facts (e.g., from external systems such as relational databases) to answer a particular goal at query time and the extensional fact base can be discarded from the memory afterwards. Overall, backward-reasoning and in particular logic programming provides a number of advantages the SLA domain. In the following we will introduce several logical extensions to the core concept of extended ContractLog LPs and derivation rules, which are needed to adequately formalize SLA rules.

### 3.2. Typed Logic and Procedural Attachments

Standard LP reasoning algorithms are derived from predicate calculus with pure formal syntactical reasoning and common rule systems and LP languages such as Prolog derivates use flat un-typed logical objects (terms) in their rule descriptions. However, SLA rules are typically defined over external business objects and business/contract vocabularies, which map to internal logical object representations in order to give SLA rules a domain-specific meaning. Moreover, service level management tools do not operate on a static internal fact base, but access a great variety of external systems such as system and network management tools. From a software engineering (SE) point of view the lack of types can been seen as a serious restriction for implementing larger rule based systems and SLA decision logics which are engineered and maintained by different people, because typical SE principles such as data abstraction and modularization are not directly supported. Therefore, rule based SLA engineering can benefit from a type concept which enables static and dynamic type checking, in order to capture the developer's intended meaning of a logic program, increase the expressiveness and the readability of the formalized SLAs, detect type errors and increase robustness and reduce the search space of goals and accordingly optimize the efficiency. *Types* in ContractLog are defined by a type relation "*t:r*", denoting that term *t* has a type *r* . ContractLog, supports two different external type systems: *Object-oriented Java class hierarchies [8]* and *Description Logic Semantic Web ontologies [18]*. The typed unification in ContractLog follows a hybrid approach and uses an external reasoner for dynamic type checking and type conversion during unification of typed or untyped terms; hence leading to a hybrid, typed logic language[18]. Before we further elaborate on the typed unification we first briefly describe both typing approaches:

**Java Types**

The object-oriented type system of Java is essentially a static type system in the sense that is does not allow parametric polymorphic type parameterization (except for generics available since Java 1.5). It supports inheritance (subclassing) and ad-hoc polymorphism with overloading and coercion (casting) as a kind of automatic type conversion between classes. In the ContractLog language the fully qualified name of the class to which a typed term/variable should belong must be used. During unification of terms ContractLog then uses the declared types, assuming the root Java type "*Object*" if no information is given (= untyped variable) and tries to unify the terms using the Java *instanceof* operator to compute subclass relations, i.e. it uses Java to do dynamic type checking. For example, a typed variable *java.lang.Integer.X* unifies with a variable *java.lang.Number.Y* since the class *Number* is a super class of *Integer*.

**Semantic Web DL Types**

Semantic Web ontology languages such as RDFS (http://www.w3.org/TR/rdf-schema/), OWL Lite or OWL DL (http://www.w3.org/2004/OWL) are syntactic XML resp. RDF mark-up counterparts of expressive, decidable description logics (DL) such as *SHIQ(D)* or *SHOIN(D)*. Description Logics are a family of KR languages that is optimized for representing the knowledge of a domain, modelling rich hierarchies of conceptual objects and instances and reasoning with these models. [19] Powerful reasoning tools such as Racer (www.sts.tu-harburg.de/~r.f.moeller/racer/), Pellet (www. mindswap.org/ 2003/ pellet/), Jena (jena. sourceforge.net/) have been developed to inference and query the DL ontologies resp. their Semantic Web counterparts. In ContractLog external RDFS or OWL ontologies describing types

Paschke, A. and Bichler, M.: Knowledge Representation Concepts for Automated SLA Management, Int. Journal of Decision Support Systems (DSS), submitted 19[th] March 2006.

(concept classes) and constant objects (individuals) can be imported to the knowledge base. Different external DL reasoners for e.g. RDFS, OWL Lite, OWL DL reasoning can be configured by the function *reasoner(<DL reasoner>)* and the defined types can be used for term typing, e.g. *X:vin_Wine* denotes that the variable *X* is of type *Wine* with *vin* being its namespace. During unification type checking is outsourced to an external reasoner (the Jena API in combination with the DL reasoner Pellet) which performs the necessary subsumption reasoning and instance inferences. The major advantage of this hybrid DL-typed approach is that existing optimized DL systems with efficient algorithms can be used for type computations.

**Operational Semantics: Typed Unification**

The operational semantics of the typed logic in ContractLog is given by a polymorphic typed unification which extends the standard untyped unification with dynamic type checking. The typed unification supports ad-hoc polymorphism where variables dynamically change their types during unification. Informally, the unification rules are defined as follows (for a formal definitions see [18]):

Untyped Unification: Ordinary untyped unification without type

Untyped-Typed Unification: The untyped query variable assumes the type of the typed target

Variable-Variable Unification:
- If the query variable is of the same type as the target variable or belongs to a subtype of the target variable, the query variable retains its type, i.e. the target variable is replaced by the query variable.
- If the query variable belongs to a super-type of the target variable, the query variable assumes the type of the target variable, i.e. the query variable is replaced by the target variable.
- If the query and the target variable are not assignable the unification fails

Variable-Constant Term Unification:
- If a variable is unified with a constant of its super-type, the unification fails
- If the type of the constant is the same or a sub-type of the variable, it succeeds and the variable becomes instantiated.

Constant-Constant Term Unification:
Both constants are equal and the type of the query constant is equal to the type of the target constant.

Complex terms such as lists are untyped by default and hence are only allowed to be unified with other untyped variables resp. variables of the highest type, i.e. "Resource" for DL types resp. "Object" for Java types. For a formal definition and a more detailed description of the typed logic in ContractLog see [18].

**Java-based Procedural Attachments**

Paschke, A. and Bichler, M.: Knowledge Representation Concepts for Automated SLA Management, Int. Journal of Decision Support Systems (DSS), submitted 19[th] March 2006.

Another extension to logic programming is *procedural attachments* which are used to dynamically call external Java-based procedural functions/methods during resolution (see [7, 8]). They enable the reuse of procedural code and facilitate the dynamic integration of facts from external data sources such as relational databases via calling their APIs using query languages such as SQL via JDBC. Java object instantiations of particular types (classes) can be bound to variables having appropriate types. During resolution the methods and attributes of the bound Java object can be used as procedural attachments within rule bodies. Static and instance methods (including calls to constructors of Java classes / objects) can be dynamically invoked (via Java reflection) taking arguments and returning a result which can possibly alter the state of the knowledge base (KB). A method that does not take arguments and does not change the KB is called an attribute. Basically, three types of procedural attachments can be used: Boolean valued attachments which can be used on the level of atoms in the rule body and object valued attachments which are treated as functions that take arguments and are executed in the context of a particular object or class in case of static method calls.

**Example 1**
add(java.lang.Integer.In1, java.lang.Integer.In2,Result):- … Result = In1 + In2
add(In1, In2, Result):- I1 = java.lang.Integer(In1), I2 = java.lang.Integer(In2),  X = I1+I2, Result = X.toString().

### 3.3. Reactive Rules: Event-Condition-Action Rules (ECA rules)

In SLA execution event-driven reactive functionalities are an obvious necessity. Typical SLA rules describe reactive decision logics following the Event-Condition-Action paradigm, e.g. "*if service is unavailable (event) and it is not maintenance time (condition) then send notification (action)*". In ContractLog we have implemented a tight integration of ECA rules into logic programming in order to represent ECA rules in a homogenous knowledge base in combination with derivation rules and facts and use the backward-reasoning rule engine as execution environment for reactive rules.

**Syntax of ECA rules in ContractLog**

The *Event-Condition-Action logic programming language (ECA-LP)* [20, 21] represents an extended ECA rule as a 6-ary function *eca(T,E,C,A,P,EL)*, where *T (time), E (event), C (condition),  A (action), P (post condition), EL(se)* are complex terms which are interpreted as queries/goals on derivation rules. The

Paschke, A. and Bichler, M.: Knowledge Representation Concepts for Automated SLA Management, Int. Journal of Decision Support Systems (DSS), submitted 19[th] March 2006.

derivation rules are used to implement the respective functionality of each of the ECA rules' parts. A complex term is a logical function of the form *c(C1,...,Cn)* with a bound number of arguments (terms) which might be constant, variable or again complex. Boolean-valued procedural attachments, as defined in section 3.2, are supported in ECA rules and can be directly used instead of a complex term. While the *E,C, A* parts of an ECA rule comply with the typical definitions of standard ECA rules (omitted here), the *T, P* and *EL* part are special extensions to ECA rules:

- The time part (T) of an ECA rule defines a pre-condition (an explicitly stated temporal event) which specifies a specific point in time at which the ECA rule should be processed by the ECA processor, either absolutely (e.g., "at 1 o'clock on the 1st of May 2006), relatively (e.g., 1 minute after event X was detected) or periodically ("e.g., "every 10 seconds").
- The post-condition (P) is evaluated after the action. It might be used to prevent backtracking from different variable bindings carrying the context information from the event or condition part via setting a cut. Or, it might be used to apply verification and validation tests using integrity constraints or test cases which must hold after the action execution, which e.g. makes an update/change of the intensional rules in the KB. If the update violates the integrity test of the post condition it is automatically rolled back.
- The else part (EL) defines an alternative action which is execute alternatively in case the ECA rule can not be applied, e.g. to specify a default action or trigger some failure handling (re-)action.

ECA parts might be left out (i.e. always true) stated with "_", e.g*., eca(time(), event(...), _, action(...),_,_)* or completely omitted, e.g. *eca(e(...),c(...),a(...))*. This leads to specific types of reactive rules, e.g. production rules (CA: *eca(condition() ,action())* ) or extended ECA rules with post condition (ECAP: *eca(event(),condition(),action(), postcondition())* ). During interpretation the smaller rule variants are extended to the full ECA rule syntax, where the omitted parts are stated as true "_".

**Operational Semantics of ECA rules in ContractLog**

In order to integrate the (re)active behavior of ECA rules into goal-driven backward-reasoning the goals defined by the complex terms in the ECA rules need to be actively used to query the knowledgebase and evaluate the derivation rules which implemented the functionality of the ECA rules' parts. Hence, an ECA rule is interpreted as a conjunction of goals (the complex terms) which must be processed in a left-to-right sequence starting with the goal denoting the time part, in order to capture the forward-directed operational semantics of ECA rules: $ECA? = T \wedge E \wedge ((C \wedge A \wedge P) \vee EL)$, where *ECA?* is the top goal and *T,E,C,A,P, EL*

Paschke, A. and Bichler, M.: Knowledge Representation Concepts for Automated SLA Management, Int. Journal of Decision Support Systems (DSS), submitted 19[th] March 2006.

are the subgoals. An ECA rule succeeds, if all subgoals succeed. Note, that the "else action" *EL* is an alternative to the normal action, which is a disjunction *(C ∧ A ∧ P) ∨ EL*. The task of interpreting ECA rules and executing the defined goals is solved by an active ECA processor with a daemon process which is build on top of the rule engine. The ECA processor implements a general wrapper interface in order to be applicable to arbitrary backward-reasoning rule engines. The daemon frequently queries the KB for new or updated ECA rules via the query *eca(T,E,C,A,P,EL)?* and adds the found ECA rules to its active KB, which is a kind of volatile storage for reactive rules. It then evaluates the ECA rules one after another via using the complex terms defined in the ECA rule as queries on the KB. The forward-directed operational semantics of ECA rules is given by the strictly positional order of the terms in the ECA rules, i.e. first the time part is queried/evaluated by the ECA processor (daemon), when it succeeds then the event part is evaluated, then the condition and so on. Figure 1 illustrates this process.

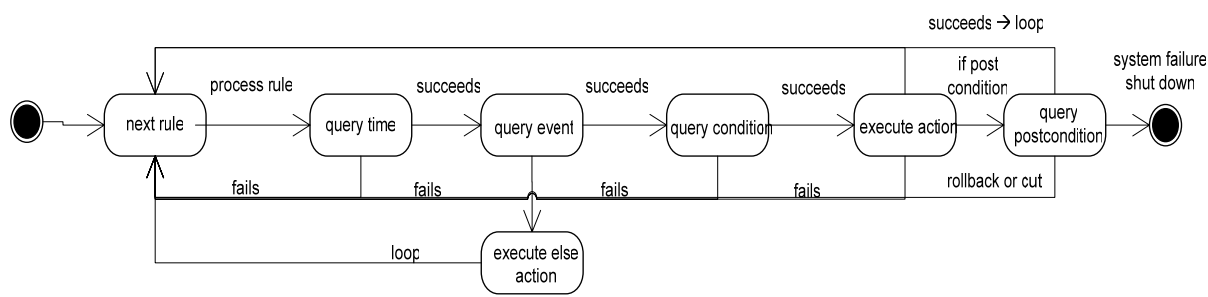

**Figure 1:** Forward-directed operational execution model of ECA rules

In order to enable parallel processing of ECA rules the ECA processor implements a thread pool where each ECA rule is executed in a thread, if its time part succeeds. Variables and negation in ECA rules are supported as described in section 3.1. For example, the ECA processor enables variable binding to ground knowledge derived by the queried derivation rules and facts in the KB, knowledge transfer from one ECA part to another via variable unification and backtracking to different variable bindings. This is in particular useful to interchange context information, e.g. between the event and the action or the condition and the action and build reactive behaviour on top of expressive knowledge bases and sophisticated derivation rule sets.

**Declarative Semantics of ECA Rules in ContractLog**

Paschke, A. and Bichler, M.: Knowledge Representation Concepts for Automated SLA Management, Int. Journal of Decision Support Systems (DSS), submitted 19[th] March 2006.

The declarative semantics of ECA rules in ContractLog LPs is directly derived from the semantics of the underlying rule/inference system. ECA rules are defined globally and unordered in the KB; homogeneously in combination with other rule types. Model-theoretically the semantics is a subset of the 3-valued (Herbrand) models of the underlying rule language: $SEM(ContractLog) \subseteq MOD_{3-val}^{Herb}(ContractLog)$ with a skeptical entailment relation $\models$, where $SEM^{skept}(ContractLog) := \bigcap_{M \in SEM(ContractLog)} \{M \models L : literal$ is the set of all atoms that are true in all models of $SEM(ContractLog)$ w.r.t. to updates of the KB (see later). Note that this general definition also covers 2-valued semantics such as Clark's completion semantics, as a declarative counterpart of the procedural "Finite-Failure SLDNF", because from a classical viewpoint the 2-valued models are special 3-valued ones, i.e. any semantics based on a two-valued theory $\Sigma$ can be considered as a three-valued interpretation $I=(T;F)$ with $T=\{A:$ a ground atom with $\Sigma \models A\}$ and $F=\{A:$ a ground atom with $\Sigma \models neg\ A\}$. Since in ContractLog the default semantics of choice is the **well-founded semantics (WFS)** resp. extended WFS (for extended LPs with default and explicit negation) by definition *SEM(ContractLog)* only contains a single model for the actual knowledge state of the KB: $SEM_{WFS}(ContractLog) = \{M\}$. Unlike other 3-valued semantics such as Fitting's semantics which are based on Kleene's strong 3-valued logic, WFS [12] is an extension of the perfect model semantics[2]. It is worth noting that WFS can be considered as an approximation of stable models for a program *P* in the sense that it coincides for a broad class of weakly stratified LPs with STABLE and classifies the same atoms as true or false as STABLE. Hence, roughly speaking, an ECA rule *eca* in an ContractLog LP *P* (or more precisely in the actual knowledge state $P_i$) is satisfied in a finite interpretation $I_{WFS}$, iff there exists at least one model for the sentence formed by the ECA top goal: $P_i \models eca$ and consequently $P_i \vdash eca \Leftrightarrow P_i \models eca$.

While this definition of soundness and completeness for ContractLog LPs is sufficient in one particular knowledge state, ECA rules typically deal with knowledge updates (see also section 3.6) which can not only be used to change the extensional knowledge (facts) but also the intensional derivation and behavioural ECA rules, hence leading to a new knowledge state. Such update actions can then be used as input events in further ECA rules leading to update sequences defined in terms of active rules. Obvi-

---
[2] hence, e.g. assigns false to a tautology such as p ← p

Paschke, A. and Bichler, M.: Knowledge Representation Concepts for Automated SLA Management, Int. Journal of Decision Support Systems (DSS), submitted 19th March 2006.

ously, updates of the intensional KB (a.k.a. dynamic LPs) might lead to confluence and logical conflicts (e.g. conflicting positive vs. negative conclusions), which are not resolved by the standard semantics of normal reps. extended LPs. Hence, this amounts for expressive update primitives with a declarative semantics for dynamically changed LPs, so called dynamic update logic programs. We first define semantics for expressive ID-based updates and unitized KB structures, where clauses are labelled by an ID (identifier or rule name) and bundled to clause sets (modules) having a module ID. A detailed description of the labelled logic in ContractLog will follow in section 3.6. The core procedural functionalities for clause sets with labels (properties) have been outlined by the original Mandarax system[3] and further extended by Prova [7] which supports dynamic imports of external LP scripts via their URL, with the URL being the module ID of the fresh module (clause set) in the KB. In the ContractLog KR we have further extended this approach with dynamic, transactional ID-based meta functions to add and remove arbitrary knowledge as modules to the KB using the module ID for partial/scoped reasoning on the clause sets, remove modules from the KB via their ID, rollback transactional updates or prioritize conflicting rules an modules (e.g., *overrides(module id1,module id2)*) (see. Section 3.6). Formally, we define a **positive (add) resp. negative (remove) ID-based update** to a program $P$ as a finite set $U_{oid}^{pos/neg} := \{rule^N : H \leftarrow B, fact^M : A \leftarrow\}$, where $N=0,..,n$, $M=0,..,m$ and *oid* denotes the label of the update, i.e. the unique object id with which it is managed in the unitized KB. Applying $U_{oid}^{pos}$ resp. $U_{oid}^{pos}$ to $P$ leads to the extended knowledge state $P^+ = P \cup U_{oid}^{pos}$ resp. reduced state $P^- = P \setminus U_{oid}^{neg}$. Applying arbitrary sequences of positive and negative updates leads to a sequence of program states $P_0,..,P_k$ where each state $P_i$ is either $P_i = P_{i-1} \cup U_{oid\ i}^{pos}$ or $P_i = P_{i-1} \setminus U_{oid\ i}^{neg}$. In other words, a program $P_i$, i.e. the set of all clauses in the KB at a particular knowledge state $i$, is decomposable in the previous knowledge state plus/minus an update, whereas the previous state consists of the state *i-2* plus/minus an update and so on. Hence, each particular knowledge state can be decomposed in the initial state $P_0$ and a sequence of updates. We define $P_0 = \emptyset \cup U_0^{pos}$ and $U_1^{pos} = \{P : the\ set\ of\ rules\ and\ facts\ defined\ in\ the\ initial\ LP\ P\}$, i.e. loading the initial LP from an external ContractLog/Prova script denotes the first update.

---

[3] http://sourceforge.net/projects/mandarax

Paschke, A. and Bichler, M.: Knowledge Representation Concepts for Automated SLA Management, Int. Journal of Decision Support Systems (DSS), submitted 19[th] March 2006.

Based on the semantics of updates we define the semantics of ECA rules in ContractLog with (self-)update actions which have an effect on the knowledge state as a **sequence of transitions** $<P,E,U> \rightarrow <P',U,U'> \rightarrow <P'',U',U''> \rightarrow .. \rightarrow <P^{n+1},U^n,A>$, where $E$ is an initiating event which triggers the update action $U$ of an ECA rule (in case the ECA condition is satisfied) and transits the initial knowledge state $P$ into $P'$. The update action $U$ might be a further sub-event in another ECA rule(s) (active rule) which triggers another update, leading to a sequence of transitions which ends with a terminating action $A$. $A$ might be e.g. an update action which is not an input event for further updates or an action without an internal effect on the knowledge state, e.g. an external notification action.

To overcome arising conflicts, preserve integrity of the KB in each state and ensure a unique declarative outcome of update sequences (active rules) or complex updates (see complex events/actions in next section) we extend the semantics of updates **to transactional updates** which are safeguarded by **integrity constraints (ICs)** or **test cases (TCs)** (see section 3.6 and [22]). ICs and TCs are used to verify and validate the actual or any hypothetically future knowledge state. [22, 23] A transactional update is an update, possibly consisting of several atomic updates, i.e. a sequence of transitions, which must be executed completely or not at all. In case a transactional update fails, i.e. it is only partially executed or violates integrity, it will be rolled back otherwise it will be committed. We define a transactional update as $U_{oid}^{trans} = U_{oid_1}^{pos/neg},...,U_{oid_n}^{pos/neg} \& C_1,..C_m$, where $C_i$ is the set of Ics (or TCs) which are used to test integrity and verify and validate the intended models of the updated clauses/modules. In short, ICs are defined as constraints on the set of possible models and describe the model(s) which should be considered as strictly conflicting. Model-theoretically the truth of an IC in a finite interpretation $I$ is determined by running the constraint definition as a goal $G_{IC}$ (proof-theoretically, by meta interpretation, as a test query) on the clauses in $P$ or more precisely on the actual state $P_i$. If the $G_{IC}$ is satisfied, i.e. there exists at least one model for $G_{IC}$: $P_i \models G_{IC}$, the IC is violated and $P_i$ is proven to be in an inconsistent state: $IC$ is violated resp. $P_i$ violates integrity iff for any interpretation $I$: $I \models P_i \rightarrow I \models G_{IC}$. For more information on the different types of ICs and the meta implementations in the ContractLog KR see section 3.6 and [4, 22, 24]. The ICs are used to test the hypothetically updated program state $P_{i+1}^{hypo} = P_i \cup U_{oid}^{trans\,pos}$ resp. $P_{i+1}^{hypo} = P_i \setminus U_{oid}^{trans\,neg}$ and rollback the (possibly partial executed) transactional update to the state $P_{i+1} = P_i$ or commit the update $P_{i+1} = P_{i+1}^{hypo}$ in case the update preserves integrity w.r.t. to the defined ICs. Hence, if there exists a satisfied IC in the hypothetical/pending state $P_{i+1}^{hypo}$ the transactional update is rolled back:

$$P_{i+1} = P_i \cup U_{oid}^{trans^{pos}} \Rightarrow P_{i+1}^{hypo} \setminus U_{oid}^{trans^{pos}} \text{ iff exists } P_{i+1} \models IC_j, j=1,..,n$$

$$P_{i+1} = P_i \setminus U_{oid}^{trans^{neg}} \Rightarrow P_{i+1}^{hypo} \cup U_{oid}^{trans^{neg}} \text{ iff exists } P_{i+1} \models IC_j, j=1,..,n$$

A commit is defined as $P_{i+1}^{hypo} \Rightarrow P_{i+1}$.

Note, that the procedural semantics for transactional updates is based on the **labelled logic** implemented in ContractLog/Prova (see section 3.6 and [7, 8, 22]), where rules and rule sets (modules) are managed in the KB as labelled clause sets having an unique oid (object identifier). Updates are new modules (or changes in existing modules) which are added to the KB, e.g. by importing new rule sets from external scripts. The modules are managed and removed from the KB by their IDs. Additional local ICs resp. TCs can be explicitly specified in the ECA post condition part and the ContractLog KR provides special functions to hypothetically tests the update literals (rule heads) against the constraints.

**Representation of Reactive Rules**

To illustrate the usage and formalization of ECA rules in ContractLog consider an ECA rule which states that: *"every 10 seconds it is checked (time) whether there is a service request by a customer (event). If there is a service request a list of all currently unloaded servers is created (condition) and the service is loaded to the first server (action). In case this action fails, the system will backtrack and try to load the service to the next server in the list. Otherwise it succeeds and further backtracking is prevented (post-condition cut) ."* This is formalized as an ECA-LP as follows:

```
eca(    every10Sec(), detect(request(Customer, Service),T),  find(Server, Service), load(Server, Service), ! ).
% time derivation rule
every10Sec() :-  sysTime(T), interval( timespan(0,0,0,10),T).
% event derivation rule
detect(request(Customer, Service),T):- occurs(request(Customer,Service),T),  consume(request(Customer,Service)).
% condition derivation rule
find(Server,Service) :- sysTime(T), holdsAt(status(Service, unloaded),T).
% action derivation rule
load(Server, Service) :- sysTime(T),
```

Paschke, A. and Bichler, M.: Knowledge Representation Concepts for Automated SLA Management, Int. Journal of Decision Support Systems (DSS), submitted 19[th] March 2006.

```
                    rbsla.utils.WebService.load(Server,Service),  % procedural attachment
                    update(key(Server), "happens(loading(_0),_1).", [Server, T]). % update KB with "loading" event
```
The state of each server might be managed via a KR event logics formalism such as the Event Calculus:

```
terminates(loading(Server),status(Server,unloaded),T).      initiates(unloading(Server),status(Server,unloaded),T).
```

### 3.4. Event/Action Logics: Event Calculus and interval-based Event/Action Algebra

Pure ECA rule processing, as described in the last section, is concerned with detecting real-time event occurrences (volatile situations) and triggering immediate reactions. But, in SLA representation there is also a need for an event/action algebra and an temporal event/action logic which is used to define complex events / actions and reason over the effects of events/actions on the knowledge state. Typical examples found in SLAs are e.g, "*After four outages then ...*", "*If the service is unavailable it must be repaired within 10 minutes. If it is still unavailable afterwards then...*" or "*If average availability is below 99% and maximum responstime is more than 4 seconds then ...*".

**Event Calculus**

Kowalski and Sergot´s Event Calculus (EC) [25] is a formalism for temporal reasoning about events/actions and their effects on LP system as a computation of earlier events (long-term "historical" perspective). It defines a model of change in which *events* happen at *time-points* and *initiate* and/or *terminate time-intervals* over which some *properties* (time-varying *fluents*) of the world hold. The basic idea is to state that *fluents* are true at particular time-points if they have been *initiated* by an event at some earlier time-point and not *terminated* by another event in the meantime. The EC embodies a notion of default persistence according to which fluents are assumed to persist until an event occurs which terminates them. In ContractLog we have implemented an optimized meta program formalization of the classical EC and extended it with a novel interval-based EC variant [20] and several other expressive features e.g. for planning, delayed effects, counters or deadlines [26, 27]. The core EC axioms describe when events/actions occur (transient view) / happen (non-transient view) / are planned (planning view) and which properties (fluents) are initiated and/or terminated by these events/actions:

| | |
|---|---|
| *occurs(E,T)* | event/action E occurs at time interval T:=[T1,T2] |
| *happens(E,T)* | event/action E happens at time T |
| *planned(E,T)* | event/action E is planned at time T |
| *initiates(E,F,T)* | event/action E initiates fluent F for all time>T |

*terminates(E,F,T)*         event/action E terminates fluent F for all time>T
*holdsAt(F,T)*              fluent F holds at time point T
*holdsInterval([E1,E2],[T1,T2])*     event/action with initiator E1 and terminator E2 holds between time T1 and T2
*holdsInterval([E1,E2],[T1,T2],[<Terminators>])* with list of terminator events which terminate the event interval [E1,E2]

**Example 2**

initiates(e1,f,T).  terminates(e2,f,T).
happens(e1,t1).happens(e2,t5).
holdsAt(f,t3)?    ➔ <u>true</u>
holdsAt(f,t7)?    ➔ <u>false</u>

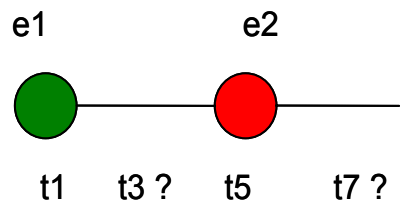

The example states that an event *e1* initiates a fluent *f* while an event *e2* terminates it. An event *e1* happens at timepoint *t1* and *e2* happens at timepoint *t5*. Accordingly a query on the fluent *f* at timepoint *t3* will succeed while it fails at timepoint *t7*. Note, that in ContractLog there is no restriction on the terms used within the EC axioms, i.e. a term can be a constant/(Java) object, a variable or even a complex term which is unified with other rules and it can be assigned a certain type dynamically at runtime (see section 3.2).

**Complex Event / Action Algebra based on the interval-based Event Calculus**

Based on the interval-based EC formalization we have implemented a *logic-based event/action algebra* [20] which provides typical operators for defining and computing complex events and actions, e.g. *sequence, conjunction, disjunction, negation*. As we have pointed out in [20] typical event algebras in the active database domain, such as Snoop [28], which considered events to be instantaneous, have unintended semantics and anomalies for several of their operators and the interval-based treatment of complex events/actions in ContractLog helps to overcome these inconsistencies and irregularities. Moreover, the formal logical basis of the used KR event logics (event calculus) in ContractLog facilitates reliable and traceable results. Using the holdsInterval axiom typical event algebra operators can be formalized in terms of the interval-based EC. For example the sequence operator ";", which defines that the specified events/actions have to occur in the specified order, can be formalized as follows: Sequence operator (;), e.g.

(a;b;c) ≡ detect(e,[T1,T3])   :- holdsInterval([a,b],[T1,T2],[a,b,c]), holdsInterval([b,c],[T2,T3],[a,b,c]), [T1,T2]<=[T2,T3].

The example defines the detection conditions (detection rules) of a complex event *e* which is defined by the two sub-event intervals *[a,b]* and *[b,c]*, where the first interval must occur before the second *[T1,T2]<=[T2,T3]*. For both sub-event intervals a list of terminator events *[a,b,c]* is provided which ter-

Paschke, A. and Bichler, M.: Knowledge Representation Concepts for Automated SLA Management, Int. Journal of Decision Support Systems (DSS), submitted 19[th] March 2006.

minate the event intervals, in order to ensure the correct sequence of events. In order to make definitions of complex events more comfortable and remove the burden of defining all interval conditions for a particular complex event type as described above, we have implemented a meta program which implements an interval-based EC event algebra with the following axioms:

| | |
|---|---|
| Sequence: | sequence(E1,E2, .., En) |
| Disjunction: | or(E1,E2, .. , En) |
| Mutual exclusive: | xor(E1,E2,..,En) |
| Conjunction: | and(E1,E2,..,En) |
| Simultaneous: | concurrent(E1,E2,..,En) |
| Negation: | neg([ET1,..,ETn],[E1,E2]) |
| Quantification: | any(n,E) |
| Aperiodic: | aperiodic(E,[E1,E2]) |

Note that, the EC makes no distinction between events and actions. Hence, the described event operators also apply for complex action definitions, e.g. a sequence of actions (e.g. sequence of knowledge updates) which must be processed in the defined order.

**Integration of Event Calculus and Complex Event/Action Algebra into ECA Rules**

The Event Calculus and the EC based event algebra can be easily integrated into ECA rules via querying the EC axioms or the complex event or action definitions. For example a SLA rule might define that *(the state) escalation level 1 is triggered (action) in case a service "s" is detected to be unavailable via a minutely (time) ping on the service (event), except we are currently in a maintenance state (condition)*. This can be formalized as an ECA rule: eca(everyMinute(), detect(unavailable(s),T), not(holdsAt( maintenance(s),T)), update("",happens( unavailable(s),T)))., i.e., in the condition part it is evaluated whether the state *maintenance* for the service *s* holds at the time of detection of the *unavailable* event or not. In case it does not hold, i.e. the condition succeeds, the detected transient event *unavailable* is added to the KB as a non-transient event in terms of a EC "happens" fact, in order to initiate the escalation level 1: initiates(unavailable(s), escl(1),T). Additionally, we might define that the *unavailable* event can not be detected again as long as the state *escl(1)* has not been terminated and accordingly the ECA rule will not fire again in this state: detect(e,T) :- not(holdsAt(escl(1),T)), ... . This exactly captures the intended behaviour for this reactive rule. The detection conditions detect(e,T) for an unavailable service might be extended by a complex event, which is detected e.g. if three consecutive pings on the service fail.

Paschke, A. and Bichler, M.: Knowledge Representation Concepts for Automated SLA Management, Int. Journal of Decision Support Systems (DSS), submitted 19[th] March 2006.

In short, the EC can be effectively used to model the effects of events on the knowledge states and describe sophisticated state transitions akin to state machines. In contrast to the original use of ECA rules in active database management systems to trigger timely response when situations of interest occur which are detected by volatile vanishing events, the integration of event logics KR formalisms adds temporal reasoning on the effects of non-transient, happened (or planned) events on the knowledge system, i.e. enable traceable "state tracking". They allow building complex decision logics upon, based on a logical semantics as opposed to the database implementations which only have an operational semantics. As a result, the derived conclusions and triggered actions become verifiable and traceable, which is a crucial necessity in SLA monitoring.

**3.5. Deontic Logic with Norm Violations and Exceptions**

One of the main objectives of a SLA is to define and reason with the normative relationships relating to permissions, obligations and prohibitions between contract partners, i.e. to define the rights and obligations each role has in particular state of the contract. Deontic Logic (DL) studies the logic of normative concepts such as obligation (O), permission (P) and prohibition (F). Adding deontic logic is therefore a useful concept for SLM tools, in particular with respect to traceability and verifiability of derived contract norms (rights and obligations). Unfortunately, standard deontic logic (SDL) offers only very a static picture of the relationships between co-existing norms and does not take into account the *effects of events on the given norms, temporal notions* and *dependencies between norms*, e.g. violations of norms or exceptions. Another limitation is the inability to express *personalized norms,* i.e. explicitly define the subject and object a norm pertains to. Therefore, we extended the general concepts of SDL and integrated it into the event calculus implementation in order to model the effects of events/actions on personalized deontic norms. A deontic norm in ContractLog consists of the normative concept (*norm*), the subject (*S*) to which the norm pertains, the object (*O*) on which the action is performed and the action (*A*) itself. We represent a deontic norm as an EC fluent of the form: *norm(S, O, A)*.

**Example 3**

initiates(unavailable(Server), escl(1),T). terminates(available(Server), escl(1),T).
initiates(maintaining(Server),status(Server,maintenance),T). terminates(maintaining(Server),escl(1),T).

```
derived(oblige(processManager, Service, restart(Service))).
holdsAt(oblige(processManager, Service, restart(Service)),T):- holdsAt(escl(1),T).
```

In the example escalation level 1 is initiated resp. terminated, when a service becomes unavailable resp. available, e.g. happens(unavailable(s1), t1) The deontic obligation for the process manager to restart the service is defined as a derived fluent, i.e. it holds whenever the state escl(1) holds. If the process manager is permitted to start maintenance (e.g. between 0 a.m. and 4 a.m. - not shown here) the second and third rule state that the event *maintaining(Server)* will initiate maintenance and terminate the escalation level 1.

The integration of deontic logic concepts into the EC enables the definition of sophisticated dependencies between events and contract norms relating to state machines. A norm can be initiated resp. terminated by an event and the EC allows inferring all actual contract state, i.e. the rights and obligations (deontic norms stated as fluents) which hold at a specific point in time according to the happened events (contract norm / state tracking). We have implemented typical SDL inference axioms in ContractLog such as $OA \rightarrow PA$ or $FA \rightarrow WA$ etc., i.e. an action "A" is permitted (P) if it is also obliged (O), it is waived (W) if it is forbidden. Moreover, the EC formalization of deontic logic avoids or overcomes typical deontic conflicts and paradoxes, i.e. sets of sentences that derive sentences with a counterintuitive reading, e.g.:

- $PA \wedge FA \rightarrow FA$, i.e. if an action *A* is permitted and forbidden, then only the prohibition holds.
- Exceptions, e.g. $OA$ and $E \rightarrow O\neg A$, i.e. *A* is obliged but in an exceptional situation *not A* is obliged
- Violations of deontic norms, e.g. $OA$ and $\neg A$, i.e. *A* is obliged but not *A* happens which is a violation
- Contrary-to-duty (CTD) obligations which hold in case of violations *V* of primary obligations $V \rightarrow OA_2$

For example a process manager is obliged to restart an unavailable service within time *t* (primary obligation). However, she fails to do so and violates the primary obligation (violation). In that case a secondary obligation (CTD) must hold, e.g. she is obliged to report this to the SLA quality manager to comply with certain compliance rules. Due to temporal framework of the EC such conflicting situation where a violated obligation and a CTD obligation of the violated obligation are true at the same time can be efficiently avoided by terminating the violated obligation so that only the consequences of the violation (CTD obligation) are in effect. [29, 30] Another way to solve such problems is to use defeasible conflict handling [4, 30].

Paschke, A. and Bichler, M.: Knowledge Representation Concepts for Automated SLA Management, Int. Journal of Decision Support Systems (DSS), submitted 19[th] March 2006.

## 3.6. Integrity Constraints, ID-based Updates and Defeasible Reasoning

Rules in SLAs might overlap and contradict each other, in particular if contracts grow larger and more complex and are authored, maintained and updated by different people.

**Example 4**

(r1) discount(Customer, 10) :- spending(Customer, Value, last year), Value > 1000.
(r2) discount(Customer, 5) :- spending(Customer, Value, last year), Value > 500.

In the example a customer might apply for a discount of "10%" as well as a discount of "5%". From an applications point of view only the higher discount should be drawn.

Nute's defeasible logic (DefL) [31] is a non-monotonic reasoning approach which allows defeasible reasoning, where the conclusion of a rule might be overturned by the effect of another rule with higher priority, i.e. it seeks to resolve conflicts by "defeating" and explicitly expressed superiority relations between rules. Defeasible logic differs between *strict rules* and *defeasible rules* (we omit defeaters here):

- Strict rules: Normal monotonic rules (derivation rules) with rule label: *r: head ⇐ body*
- Defeasible rules are rules that can be defeated by contrary rules: *r: head <= body*
- Priority relations are used to define priorities among rules to represent that one rule may override the conclusion of another (defeasible) rule: *r1 > r2*

Different variants have been proposed, reaching from simple defeasible implementations which deal with conflicts between positive and negative conclusions [32] to Generalized Courteous Logic Programs (GCLP) [33] which use an additional "Mutex" (mutual exclusive) to define and handle arbitrary mutual exclusive literals. Several meta programming approaches have been proposed [34] to execute a defeasible theory in a logic program. In ContractLog we have generalized the basic concept of defeasible reasoning and combined it with the concept of integrity constraints and labelled modules (rules sets with IDs) using a meta program approach. An integrity constraint expresses a condition which must always hold. In ContractLog we support four basic types of integrity constraints: [4, 22]

1. *Not-constraints*: express that none of the stated conclusions should be drawn: integrity( not( $p_1$(…), .. , $p_n$(…))).
2. *Xor-constraints*: express that the stated conclusions are mutual exclusive, i.e. should not be drawn at the same time: integrity(xor($p_1$(…),..,$p_n$(…))).
3. *Or-constraints*: express that at least one of the stated conclusions should be drawn: integrity(or($p_1$(…),..,$p_n$(…))).
4. *And-constraints*: express that all of the stated conclusion should be drawn: integrity (and($p_1$(…), .. , $p_n$(…))).

Integrity constraints might be conditional, i.e. stated as integrity rules of the form: integrity(…):- $b_1$(…)..$b_m$(…). where the integrity constraint is valid, if all prerequisites defined in the rules body hold. We

Paschke, A. and Bichler, M.: Knowledge Representation Concepts for Automated SLA Management, Int. Journal of Decision Support Systems (DSS), submitted 19[th] March 2006.

have implemented a meta programming approach in ContractLog [4, 22] which is used to test the integrity constraints specified within an ContractLog LP. The core axioms are:

1. **testIntegrity()** enumerates all integrity constraints and tests them based on the actual facts and rules in the knowledge base
2. **testIntegrity(Literal)** tests the integrity of the LP extended with the literal, i.e. it makes a test of the hypothetically added/removed literal, which might be a fact or the head of a rule.

**Example 5**

integrity(xor(discount(C,5), discount(C,10))).

testIntegrity(discount("Adrian",X))?   %query test integrity

The example defines an integrity constraint, which states that a discount of 5% and a discount of 10% for the customer are mutually exclusive, i.e. are not allowed to occur at the same time.

Integrity constraints might be used for example to hypothetically test knowledge updates before they are applied/committed. In ContractLog we have implemented support for expressive *(transactional) ID based updates* [20, 22] which facilitate bundling of rule sets to modules including imports of external rule modules/scripts. Each module (rule set) has a unique ID with which it can be added or removed from the KB.

**Example 6**

```
update("./examples/test/test.prova").            % add an external script
update(id1,"r(1):-f(1). f(1).").                 % add rule "r(1):-f(1)." and fact "f(1)." with ID "id1"
update(id2,"r(X):-f(X).").                       % add rule "r(X):-f(X)." with ID "id2"
p(X,Y) :- update(id3, "r(_0):-f(_0), g(_0). f(_0). g(_1).", [X,Y]).    %Object place holders _N: _0=X ; _1=Y.
remove(id1).                                     % remove update/module with ID "id1"
remove("./examples/test/test.prova").            % remove external update
```

Transactional updates *transaction(update(...))* make an additional test on all integrity constraints defined in the KB or an explicitly stated integrity constraint. If the tests fail the update is rolled back. For a definition of the semantics of integrity constraints, labelled rules and ID-based updates see [22].

Based on the integrity constraints we have generalized the concept of defeasible reasoning, i.e. a conflict might not be just between positive and negative conclusions as in standard defeasible theories but also between arbitrary opposing conclusions, which are formalized as integrity constraints. Priorities might be defined between single rules but also between complete rule sets (modules) enabling rule set alternatives and hierarchical module structures (contract modules). We translate this generalized defeasible theory $D$ into a logic program $P(D)$ as follows:

Paschke, A. and Bichler, M.: Knowledge Representation Concepts for Automated SLA Management, Int. Journal of Decision Support Systems (DSS), submitted 19[th] March 2006.

(1) We define that strict knowledge is also defeasibe provable with a general rule in our meta program: defeasible([P|Args]):-bound(P), derive([P|Args]). The second-order "derive" predicate allows calling a predicate dynamically with the predicate symbol unknown until run-time, i.e. whenever *[P|Args]* which is translated into predicate notation $P(Args_1,...,Args_N)$ can be evaluated this predicate is also defeasible provable.

(2) Each priority relation $r_1>r_2$ where $r_l$ are the rule names (rule object identifiers) is stated as: overrides(r1,r2). Priorities might be also defined between rule modules: overrides(moduleOID1, moduleOID2).

(3) Each defeasible rule *r: p <= $q_1,..,q_n$* is translated into the following set of meta rules:

oid(m,p). % relation between rule head and module
oid(r,p). % relation between rule head and rule label
neg(blocked(defeasible(p))):- testIntegrity(p), defeasible($q_i$).    % used in meta program for reasoning
body(defeasible(p)) :- defeasible($q_i$). % used in meta program for reasoning
defeasible(p) :-                  % defeasible rule
    testIntegrity(p), % strictly overtuned ?
    defeasible($q_i$), % body holds defeasible?
    not(defeated(r,p)). % not defeasible defeated?

- The defeasible rule name *r* is related to the rule head *p* (where *p* is $p(x_1,...,x_k)$) and related to the module (set of rules) *m* via the predicate *oid*.
- The defeasible rule *p* is not blocked (*neg(blocked())*) if the defeasible rule does not violate the integrity of any strict knowledge (normal strict rule or fact) in the knowledge base and all prerequistes $q_i$ for all $i \in \{1,..n\}$ are defeasible provable.
- The defeasible rule *defeasbile(p)* is provable if it does not violate the integrity of strict knowledge and all prerequisites $q_i$ defeasibly hold, and it is not defeated by any other defeasible rule.

The defeasible meta program implemented in ContractLog tests whether the defeasible rule *r* with the head *p* is defeated by testing if the rule defeasibly violates the integrity of the logic program according to the defined integrity constraints. For each integrity constraint where *p* is a member the defeasible integrity test meta program does the following:

1. Test whether all conflicting literals of *p* are blocked, i.e. can not be derived using the *neg(blocked(…))* rules:    neg(blocked(defeasible(Opposer)))
2. If a conflicting literal is not blocked, test whether it is overridden by the defeasible rule *p* or by the module *p* belongs to, i.e. has higher priority than the opposer: overrides(Rule,Opposer). and overrides(RuleModuleOID,OpposerModuleOID).

If a conflicting defeasible rule, i.e. a rule which is defined in an integrity constraint to be conflicting, is not blocked and is of higher priority than the defeasible rule *p*, *p* is defeated and will be not concluded. Supperiority relations between defeasible rules can be either defined based on the rule names (rule oids) such as

overrides(r1,r2) or based on head literals such as overrides(p1(X),p2(X)), i.e. a rule (or fact) with head *p1(X)* overrides a rule with head *p2(X)*. This is very useful to define general superiority rules such as "prefer positive knowledge over negative knowledge": overrides([P|Args], neg([P|Args])) :- bound(P).

**3.8. Summary**

Contractual logic in SLAs requires several logic formalisms in order to adequately represent respective rules. Such a KR needs to be expressive, but at the same time computationally efficient. Several expressive logical extensions to the basic theories have been presented and we have illustrated how these extended KR concepts can be represented effectively as (meta) logic programs which can be executed in generic LP rule engines. Table 8 provides a short summary of the ContractLog KR framework.

Table 7: Main logic concepts integrated in ContractLog

| Logic | Formalism | Usage |
|---|---|---|
| Extended Logic (section 3.1) | Derivation Rules and extended LPs | Deductive reasoning on SLA rules extended with negation-as-finite-failure and explicit negation. |
| Typed Logic (section 3.2) | Object-oriented Typed Logic and Procedural Attachments | Typed terms restrict the search space and enable object-oriented software engineering principles. Procedural attachments integrate object oriented programming into declarative rules. → integration of external systems |
| Description Logic (section 3.2) | Hybrid Description Logic Types and Semantic Web Ontology Languages | Semantic domain descriptions (e.g., contract ontologies) in order to describe rules domain-independent. → external contract/domain vocabularies |
| (Re)active Logic (section 3.3) | Extended Event-Condition-Action Rules (ECA) | Active event detection/event processing and event-triggered actions. → reactive rules. |
| Temporal Event/Action Logic (section 3.4) | Event Calculus | Temporal reasoning about dynamic systems, e.g. interval-based complex event definitions (event algebra) or effects of events on the contract state → contract state tracking → reasoning about events/actions and their effects |
| Deontic Logic (section 3.5) | Deontic Logic with norm violations and exceptions | Rights and obligations formalized as deontic contract norms with norm violations (contrary-to-duty obligations) and exceptions (conditional. defeasible obligations). → normative deontic rules. |
| Integrity Preserving, Preferenced, Defeasible Logic (section 3.6) | Defeasible Logic and Integrity Constraints | Default rules and priority relations of rules. Facilitates conflict detection and resolution as well as revision/updating and modularity of rules. → default rules and rule priorities |
| Test Logic [22] | Test-driven Verification and Validation for Rule Bases | Validation and Verification of SLA specifications against predefined SLA requirements → safeguards the engineering, dynamic adaption and interchange process of SLAs |

## 4. Implementation: Rule Based SLA-Management

Based on the ContractLog concepts described above, we have implemented the RBSLM tool as a proof of concept implementation. Figure 2 shows the general architecture of the **r**ule *based service level management tool (RBSLM)*.

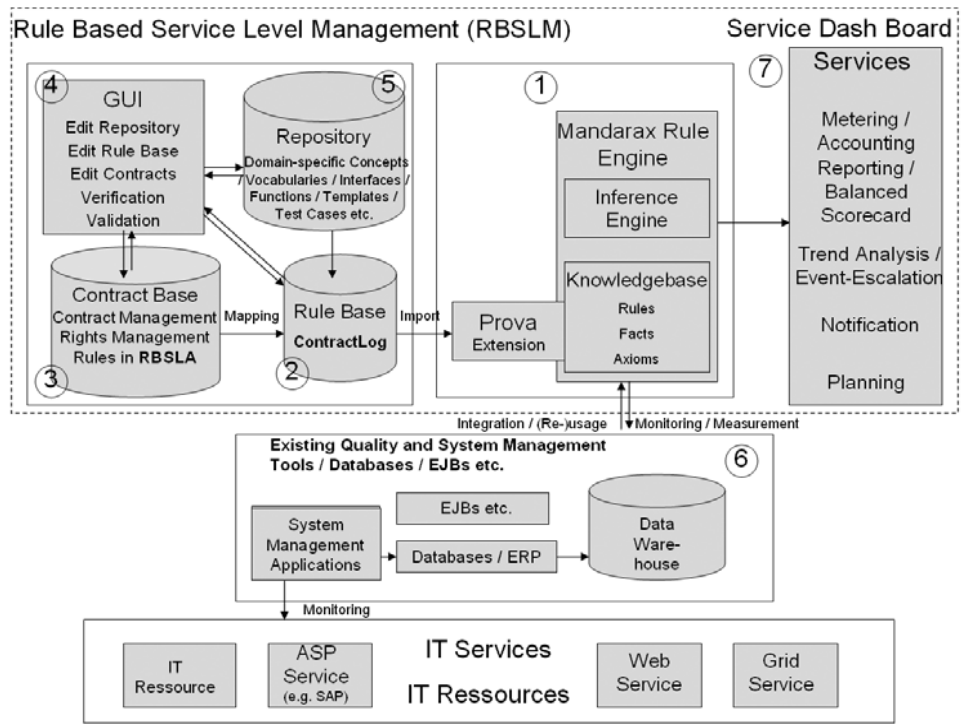

Fig. 2: Architecture of Rule Based Service Level Management Tool

The open-source, backward-reasoning rule engine Mandarax (1) (http://mandarax.sourceforge.net/) with the Prova scripting language extension (http://comas.soi.city.ac.uk /prova/), which we further extended with goal memoization, efficient data structures and different resolution algorithms and semantics (see section 3.1) serves as inference engine for the logical contract rules. The rules are represented on the basis of the *ContractLog* framework (2) and are imported using the *Prova scripting language* [7] into the internal knowledgebase of the rule engine. A high-level declarative mark-up language called the *rule based SLA (RBSLA)* language [35, 36] is provided to facilitate rule interchange, serialisation, tool based editing and verification of rules. A XSLT based mapping is defined which transforms RBSLA into executable ContractLog rules. The graphical user interface - the *Contract Manager* (4) - is used to write, edit and maintain the SLAs which are persistently stored in the contract base (3). The *repository* (5) contains typical rule templates and predefined SLA domain specific objects, built-in metrics and contract vocabularies (ontologies) which can be reused in the SLA specifications. During the enforcement and monitoring of the

Paschke, A. and Bichler, M.: Knowledge Representation Concepts for Automated SLA Management, Int. Journal of Decision Support Systems (DSS), submitted 19th March 2006.

SLAs external system management tools and business objects can be integrated (6) (via procedural attachments). Finally, the *Service Dash Board* (7) visualizes the monitoring results and supports further SLM processes, e.g. reports on violated services, metering and accounting functions, notification services.

## 5. Evaluation

ContractLog provides compact, declarative knowledge representation for contractual agreements based on logic programming. On the one hand, this enables high levels of flexibility and easy maintenance as compared to traditional SLA approaches. On the other hand, generic rule engines allow for an efficient execution of SLAs. In the following, we evaluate our KR approach by means of experiments and on an example derived from common industry use cases. For a formal analysis of the average and worst case complexity of LPs see [37].

### 5.1 Experimental Performance Evaluation

To experimentally benchmark performance of the ContractLogs' formalisms w.r.t. query answering in different logic classes (e.g. propositional, Datalog, normal) we adapt a benchmark test suite for defeasible theories [34] to the ContractLog KR and extend it to evaluate different inference properties, rule types and logic program classes. The experiments are designed to test different performance/scalability aspects of the inference engine and the logical formalisms. We ran the performance tests on an Intel Pentium 1,2 GHz PC with 512 MB RAM running Windows XP. We use different benchmark tests to measure the time required for proofs, i.e. we test performance (time in CPU seconds to answer a query) and scalability (size of test in number of literals). Various metrics such as number of facts, number of rules, overall size of literal indicating the size of complexity of a particular benchmark test might be used to estimate the time for query answering and memory consumption. The first group of benchmarks evaluates different inference aspects of the ContractLog KR such as rule chaining, recursion, and unification. In the *chains test* a chain of $n$ rules and one fact at its end is queried. In the *dag test* a directed-acyclic tree of depth $n$ is spanned by the rules in which every literal occurs recursively $k$ times. In the tree test a fact is at the root of a k-branching tree of depth $n$ in which every literal occurs once. Each test class is performed for propositional LPs without variables and Datalog LPs with one variable. The tests further distinguish strict LPs, i.e.

normal LPs with only strict (normal) derivation rules and defeasible LPs with defeasible derivation rules and compare inference for both classes with goal memoization and without goal memoization. We use as a measure of problem size the total number of literals, which is accordingly much larger for the defeasible theories due to the meta program transformations, as described in section 3.6. The second and third group of experiments test the ECA processor for (re)active rule processing and complex event processing based on Event Calculus formulations. Table 8 shows an extract of important performance results in CPU seconds. [37]

TABELLE 8: PERFORMANCE EVALUATION

| Test | Size | | Time (in seconds) | | | |
| --- | --- | --- | --- | --- | --- | --- |
| | | | No Memoization | | Memoization | |
| | Strict | Defeasible | Strict (Propos. / Datalog) | Defeasible (Propositional / Datalog) | Strict (Propositional / Datalog) | Defeasible (Propos. / Datalog) |
| $chains(n)$ | 2001 | 11001 | 0.01 / 0.07 | 4 / 7.6 | 0.05 / 0.17 | 5,7 / 7,8 |
| | 5001 | 27501 | 0.03 / 0.17 | 12.8 / 25 | 0,15 / 0.47 | 18 / 24,3 |
| | 10001 | 55001 | 0.07 / 0.3 | 40 / 70 | 0,4 / 1.05 | 59 / 75 |
| | 20001 | 110001 | 0.15 / 0.62 | 127 / 250 | 1,25 / 2,62 | 170 / 200 |
| $dag(n,k)$ | | | | | | |
| $n=3\ k=3$ | 39 | 156 | 0.01 / 0.06 | 0.54 / 0.89 | 0.005 / 0.01 | 0.05 / 0.05 |
| $n=4\ k=4$ | 84 | 324 | 2.2 / 7.7 | 81 / 120 | 0.01 / 0.03 | 0.06 / 0.07 |
| $n=10\ k=10$ | 1110 | 3810 | - / - | - / - | 0.05 / 0.16 | 0.2 / 0.32 |
| $tree(n,k)$ | | | | | | |
| $n=3\ k=3$ | 79 | 248 | 0.01 / 0.02 | 0.04 / 0.04 | 0.001 / 0.001 | 0.04 / 0.05 |
| $n=4\ k=3$ | 281 | 761 | 0.015 / 0.03 | 0.09 / 0.1 | 0.005 / 0.006 | 0.08 / 0.11 |
| $n=8\ k=3$ | 19681 | 62321 | 0.17 / 0.5 | - / - | 0.02 / 0.04 | 0.09 / 0.14 |
| | | | Update Time | | Execution Time | |
| $eca_{plain}(n)$ | 1000 | | 0.4 | | 0.005 | |
| | 2500 | | 1.1 | | 0.01 | |
| | 5000 | | 2.5 | | 0.015 | |
| | 10000 | | 4.3 | | 0.02 | |
| $ec_{holdsAt}(n)$ | 1002 | | 3.3 | | | |
| | 2502 | | 6.8 | | | |
| | 5002 | | 14.6 | | | |
| | 10002 | | 28.7 | | | |

"-" denotes that the experiment was exhausted due to excessive runtime or memory requirements

Each experiment was performed several times (10 experiments per benchmark test) and table 8 shows the average results. The table shows the computation time to find an answer for a query. Due to the needed variable unifications and variable substitutions in the derivation trees the datalog tests are in general a little bit more expensive than the propositional tests and goal memoization adds small extra costs to rule derivations. In the chains test where subgoals are never reused the experiments with memoization are slower than without memoization due to the caching overhead. However, the advantages of goal memoization can be seen in the tree and dag tests which (recursively) reuse subgoals. Here goal memoization leads to much higher performance and scalability (large problem sizes can be solved). As expected the defeasible experiments run slower, due to the much larger problem sizes and the meta program interpretation which need several KB subgoal queries. Goal memoization reduces duplication of work, e.g. to test strict integrity of defeasible rules. The ECA rule experiments distinguish between update time for querying the KB for ECA rules and processing time for executing the ECA rules. The experiments reveal an increase in time linear in the problem size. The event calculus test also show linear time increase in the problem size, which here is the number of occurred events stated as happens facts which initiate resp. terminate a fluent.

In summary, the experiments reveal high performance of the ContractLog formalisms even for larger problem sizes with thousands of rules and more than 10000 literals, which qualifies the approach also for industrial applications. We have formalized typical real-world SLAs from different industries in ContractLog within several dozens up to hundreds rules and much smaller literal sizes (see e.g. RIF / RuleML use cases[4]) which can be efficiently executed and monitored within milliseconds. Moreover, the hybrid approach in ContractLog allows outsourcing lower-level computations and operational functionalities to procedural code and specialized external systems (e.g. DL reasoner [18]).

**5.2. Use Case Revisited – Adequacy / Expressiveness**

In this section we illustrate adequacy of the rule-based SLA representation approach, in particular with respect to expressiveness of the ContractLog KR, by means of a use case example derived from common industry SLAs. We revive the example SLA described in section 2.2 and present a formalization of a selected subset in ContractLog, namely the monitoring schedules, the escalation levels and the associated roles, as well as the following SLA rules:

**Example 7:** ContractLog formalization of SLA

*"The service availability will be measured every $t_{schedule}$ according to the actual schedule by a ping on the service. If the service is unavailable and it is not maintenance then escalation level 1 is triggered and the process manager is informed. Between 0-4 a.m. the process manager is permitted to start servicing which terminates any escalation level. The process manager is obliged to restart the service within time-to-repair, if the service is unavailable. If the process manager fails to restore the service in time-to-repair (violation of obligation), escalation level 2 is triggered and the chief quality manager is informed. The chief quality manager is permitted to extend the time-to-repair interval up to a defined maximum value in order to enable the process manager to restart the service within this new time-to-repair. If the process manager fails to restart the service within a maximum time-to-repair escalation level 3 is triggered and*

---

[4] http://www.w3.org/2005/rules/wg/wiki/Rule_Based_Service_Level_Management_and_SLAs_for_Service_Oriented_Computing

Paschke, A. and Bichler, M.: Knowledge Representation Concepts for Automated SLA Management, Int. Journal of Decision Support Systems (DSS), submitted 19th March 2006.

*the control committee is informed. In escalation level 3 the service consumer is permitted to cancel the contract."*

The formalization in ContractLog is as follows:

```
% service definition
service(http://ibis.in.tum.de/staff/paschke/rbsla/index.htm).
% role model and escalation levels
initially(escl_lvl(0)).         % initially escalation level 0
role(process_manager) :- holdsAt(escl_lvl(1),T). % if escalation level 1 then process_manager
role(chief_quality_manager) :-  holdsAt(escl_lvl(2),T). % if escalation level 2 then chief quality manager
role(control_committee) :- holdsAt(escl_lvl(3),T). % if escalation level 3 then control committee
% time schedules standard, prime, maintenance and monitoring intervals
% before 8 and after 18 every minute
schedule(standard, Service):-
        systime(datetime(Y,M,D,H,Min,S)), less(datetime(Y,M,D,H,Min,S), datetime(Y,M,D,8,0,0)),
        interval(timespan(0,0,1,0), datetime(Y,M,D,H,Min,S)),
        service(Service), not(maintenance(Service)).  % not maintenance
schedule(standard, Service):-
        systime(datetime(Y,M,D,H,Min,S)), more(datetime(Y,M,D,H,Min,S), datetime(Y,M,D,18,0,0)),
        interval(timespan(0,0,1,0), datetime(Y,M,D,H,Min,S)),
        service(Service), not(maintenance(Service)).  % not maintenance
% between 8 and 18 every 10 seconds
schedule(prime, Service):-
        sysTime(datetime(Y,M,D,H,Min,S)), lessequ(datetime(Y,M,D,H,Min,S),datetime(Y,M,D,18,0,0)),
        moreequ(datetime(Y,M,D,H,Min,S),datetime(Y,M,D,8,0,0)),
         interval(timespan(0,0,0,10), datetime(Y,M,D,H,Min,S)) , service(Service).
% between 0 and 4 if maintenance every 10 minutes
schedule(maintenance, Service) :-
        sysTime(datetime(Y,M,D,H,Min,S)), lessequ(datetime(Y,M,D,H,Min,S),datetime(Y,M,D,4,0,0)),
        interval(timespan(0,0,10,0), datetime(Y,M,D,H,Min,S)) ,
        service(Service), maintenance(Service). % servicing
initiates(startServicing(S),maintenance(S),T). % initiate maintenance if permitted
terminates(stopServicing(S), maintenance(S),T). % terminate maintenance
happens(startServicing(Service),T):-
        happens(requestServicing(Role,Service),T), holdsAt(permit(Role,Service, startServicing(Service)),T).
% ECA rule: "If the ping on the service fails and not maintenance then trigger escalation level 1 and notify
process manager, else if ping succeeds and service is down then update with restart information and inform
responsible role about restart".
eca(schedule(T,S), not(available(S)), not(maintenance(S)), escalate(S),_, restart(S)). % ECA rule

available(S) :- WebService.ping(S).           % ping service
maintenance(S) :- sysTime(T), holdsAt(maintenance(S),T).
escalate(S) :-    sysTime(T), not(holdsAt(unavailable(S),T)), % escalate only once
                  update("outages","happens(outage(_0),_1).",[S,T]),% add event
                  role(R), notify (R, unavailable(S)).  % notify
restart(S) :- sysTime(T), holdsAt(unavailable(S),T), update("outages","happens(restart(_0),_1).",[S,T]),% add event
          role(R), notify(R,restart(S)).       % update + notify
% initiate unavailable state if outage event happens
initiates(outage(S),unavailable(S),T).            terminates(restart(S),unavailable(S),T).
% initiate escalation level 1 if outage event happens
terminates(outage(S),escl_lvl(0),T). initiates(outage(S),escl_lvl(1),T).
% terminate escalation level 1/2/3 if restart event happens
initiates(restart(S),escl_lvl(0),T). terminates(restart(S),escl_lvl(1),T). terminates(restart(S),escl_lvl(2),T). termi-
nates(restart(S), escl_lvl(3), T).
% terminate escalation level 1/2/3 if servicing is started
```

Paschke, A. and Bichler, M.: Knowledge Representation Concepts for Automated SLA Management, Int. Journal of Decision Support Systems (DSS), submitted 19[th] March 2006.

```
initiates(startServicing(S),escl_lvl(0),T). terminates(startServicing(S), escl_lvl(1),T). terminates(startServicing(S),
    escl_lvl(2),T). terminates(startServicing(S),escl_lvl(3),T).
% permit process manager to start servicing between 0-4 a.m.
holdsAt(permit(process_manager,Service, startServicing(Service)), datetime(Y,M,D,H,Min,S)):-
        lessequ(datetime(Y,M,D,H,Min,S),datetime(Y,M,D,4,0,0)).
% else forbid process manager to start servicing.
holdsAt(forbid(process_manager,Service, startServicing(Service)), datetime(Y,M,D,H,Min,S)):-
        more(datetime(Y,M,D,H,Min,S),datetime(Y,M,D,4,0,0))..
% derive obligation to start the service if service unavailable
derived(oblige(process_manager, Service , restart(Service))). % oblige process manager
holdsAt(oblige(process_manager, Service , restart(Service)), T) :- holdsAt(unavailable(Service),T).
% define time-to-repair deadline and trigger escalation level 2 if deadline is elapsed
time_to_repair(t_deadline). % relative time to repair value
trajectory(escl_lvl(1),T1,deadline,T2,(T2 - T1)) .          % deadline function
derivedEvent(elapsed).
happens(elapsed,T) :- time_to_repair(TTR),  valueAt(deadline,T, TTR).
terminates(elapsed, escl_lvl(1),T).% terminate escalation level 1
initiates(elapsed, escl_lvl(2),T). % initiate escalation level 2
% trigger escalation level 3 if (updated) time-to-repair is > max time-to-repair
happens(exceeded,T) :- happens(elapsed,T1), T=T1+ ttr_max.
terminates(exceeded,escl_lvl(2),T). initiates(exceeded, escl_lvl(3),T).
% service consumer is permitted to cancel the contract in escl_lvl3
derived(permit(service_consumer, contract , cancel)).
holdsAt(permit(service_consumer, contract , cancel), T) :- holdsAt(escl_lvl(3),T).
```

Via simple queries the actual escalation level and the rights and obligations each role has in a particular state can be derived from the rule base and the maximum validity interval (MVI) for each contract state, e.g. the maximum outage time, can be computed. These MVIs can be used to compute the service levels such as average availability. The ECA processor of the ContractLog framework actively monitors the ECA rules. Every $t_{check}$ according to the actual schedule it pings the service via a procedural attachment, triggers the next escalation level if the service is unavailable and informs the corresponding role.

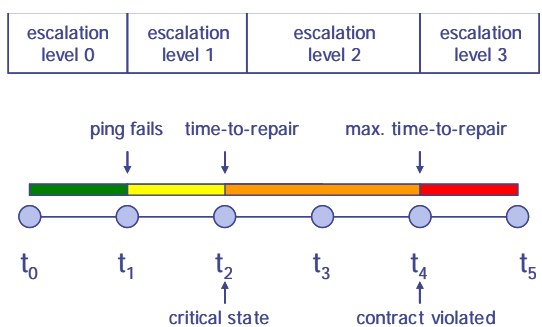

**Fig. 3**: **Contract tracking**

To illustrate this process, we assume that the service becomes unavailable at time *t1*. Accordingly, escalation level 1 is triggered and the process manager has time-to-repair *t2*. After *t2* escalation level 2 is triggered and the chief quality manager adapts the time-to-repair to *t3* and then to *t4* until the maximum threshold max. time-to-repair is reached at time point *t4*. After *t4* the SLA is violated and escalation level 3 is initiated which permits the service consumer to terminate the contract. Via querying the rule engine these status information can be dynamically derived at each point in time and used to feed periodical re-ports, enforce rights and obligations or visualize monitoring results on quality aspects in the Service Dashboard (see figure 3).

As can be seen from this use case example, the declarative rule based approach allows a very compact representation of globally defined SLA rules, which would not be possible in standard imperative programming languages. To encapsulate this QoS monitoring and decision making logic in Java an enormous object oriented infrastructure with several Java classes and multiple methods to model all subtle nuances would be needed. The entire control flow must be specified, i.e. the exact order and number of steps and decisions would be implicitly encoded into the procedural code. Obviously, this does not always scale and maintenance and management of the knowledge structures becomes increasingly difficult, especially when the SLA logic is likely to change frequently. In contrast to if-then structures in Java which form a hierarchical tree, logical rules assemble something more like a complex net, with dozens or even thousands of interconnected global rule nodes. Hence, the declarative approach naturally supports reuse of rules by global visibility. While in Java the if-then statements need to be processed from the beginning to the end exactly in the predefined order, in the declarative approach the derivation net can be queried at any node and the inference engine solves the task of finding all specific knowledge out of the general rules which assert general information about a SLA decision making problem. This global, declarative approach provides much flexibility when adapting SLA rules to changing needs, for example, when exceptional discounting policies are added.. This ability to dynamically and quickly alter the SLA logic at runtime without any extensions to the generic inference engine is a key advantage, which would require reimplementation of the application code resp. database schemas and perhaps larger service execution outages for redeployment.

Hence, by representing the SLA monitoring and enforcement logic on a more abstract rule-based level and separating it form the low-level procedural aspects of service runtime environment, much more powerful SLA specifications can be written and maintenance and management of large numbers of complex individualized SLAs becomes much easier. The clear mathematical basis of the selected logical formalisms ensures correctness and traceability of derived results, which is crucial in the SLA domain in order to provide reliable and provable reactions and results, e.g. computed penalties used in accounting. Furthermore, it enables easier validation and verification of the SLA specifications and therefore ensures consistency (due to sound and complete logical semantics) and integrity (integrity constraints / test cases) as well automated conflict resolution (via defeasible refutation).

This section should provide some evidence that the proposed ContractLog framework provides KR adequacy criteria such as *epistomological adequacy* ("are we able to represent all relevant facts and rules?"), and *heuristical / algorithmic / logic-formal adequacy* ("are we able to execute all inferences with the limited resources?" and "Is the logical system sound, complete and decidable?" ).

## 6. Related Work

The tendency in common commercial SLA/SLM tools such as IBM Tivoli SLA, HP OpenView, CA Unicenter, BMC Remedy Service Management is to allow specification of QoS parameters (e.g. availability, response time) with high/low bounds. Typically, these parameters are directly encoded in the application code or database tier, which makes it complicated to dynamically extend the SLA logic and describe more complex specifications than simple bounds. Hence, this approach is restricted to simple, static rules with only a limited set of parameters. More complex conditionals where one parameter depends upon some other parameters / conditions are not expressible. For instance a certain response time level may be necessary during prime time, a lower availability might be necessary under higher server loads or at least response time must be high when availability falls below a certain value. Due to the implicit procedural encoding of the SLA logic into the application code the SLAs are hard to manage, maintain and in particular adapt to new requirements, which would require heavy time and cost intensive refactorings of the application code and database schemas.

There are several XML based mark-up approaches towards specification and management languages for SLAs such as the IBM Web Service Level Agreements (WSLA) [38], the HP Web Service Management Language (WSML) [39] or the Web Service Offering Language (WSOL) [40]. These languages include definitions of the involved parties (signatory and supporting parties), references to the operational service descriptions (e.g. WSDL) of the service(s) covered by the SLA, the SLA parameters to be monitored and the metrics and algorithms used to compute SLA parameters from raw metrics collected by measurement directives from external sources. They allow the specification of QoS guarantees, constraints imposed on SLA parameters and compensating activities in case these constraints are violated, in terms of implica-tional clauses with simple Boolean evaluation functions and explicit Boolean connectives. Accordingly, these rules only have a very limited expressiveness restricted to truth/false implications without variables, quantifications, rule chaining and (non-monotonic) reasoning inferences. Hence these languages can not express more than simple conditional clauses and the interpreted must be adapted each time new functionalities are introduced into the serialization language.

There have been some proposals on using Petri-nets [41] or Finite State Machines [42] to describe contracts as process flows. However, those approaches are best suited for contracts which follow a pre-defined protocol sequence. Recently, in the area of policy specifications there are several proposals which address the definition of policies such as WS-Policy [43] or WS-Agreement, or policy languages such as KAos, Rei or Ponder . However, the former are only general frameworks, where the details of special policy aspects still need to be defined in specialized sublanguages, and the later mainly focus on typical operational policies such as access control or security issues and only require/consider a very limited set of logical formalisms. In particular they are missing expressive rules (reactive rules, defeasible rules, normative rules).

To our knowledge the only work which deals with contractual agreements in a declarative logical context is the work of Grosof et al. [44] The Semantic Web Enabling Technology (SWEET) toolkit comprises the CommonRules syntax and also enables business rules to be represented in RuleML. Whilst their approach deals with contracts in a broader range namely e-commerce contracts and mainly supports rule priorities via Generalized Courteous Logic Programs (GCLP) [33] and to some extend procedural attachments, our approach is focused on the specifics of Service Level Management. In contrast to SWEET, ContractLog incorporates additional logical concepts which are needed for adequate SLA representation such as contract states, explicit rights and obligations (deontic norms) supplemented with violations and exceptions of norms, integrity constraints, event processing facilities with active sensing, monitoring and triggering of actions, full support for different type systems (e.g. Java, Semantic Web) and procedural attachments in order to integrate existing business object implementations, and SLA-specific contract vocabularies.

Paschke, A. and Bichler, M.: Knowledge Representation Concepts for Automated SLA Management, Int. Journal of Decision Support Systems (DSS), submitted 19[th] March 2006.

## 7. Conclusions

Logic programming has been a very popular paradigm in the late 1980's and one of the most successful representatives of declarative programming in general. Although, logic programming is based on solid and well-understood theoretical concepts and has been proven to be very useful for rapid prototyping and describing problems on a high abstraction level, its application in commercial software has been limited throughout the past years. Service level management and more generally contract management appear to be particularly suitable to logic programming. IT service providers need to manage large amounts of SLAs with complex contractual rules describing various decision and business logic, reaching from deeply nested conditional clauses, reactive or even proactive behaviour to normative statements and integrity definitions. These rules are typically not of static nature and need to be continuously adapted to changing needs. Furthermore, the derived conclusions and results need to be highly reliable and traceable to count even in the legal sense of a contract. This demands for a declarative knowledge representation language which is computationally efficient even for larger SLA specifications, reliable and traceable even in case of incomplete or contradicting knowledge, flexible in a way that allows to quickly alter the behaviour of the SLA system and supports *programming* of arbitrary functionalities and decision procedures. Extended logic programs and derivation rules have several advantages over imperative languages such as Java, or database solutions. However, general logic programs need to be extended by multiple knowledge representation concepts and integrated with commercial system management tools to allow formalising the inherent complexity of SLAs and provide the necessary basis for service level management.

In this article we have described ContractLog, a framework of integrated knowledge representation concepts to define and automatically enforce large amounts of SLAs based on generic derivation rule engines. In contrast to a conventional procedural implementation or pure formal specification approaches the declarative, rule-based (programming) approach in ContractLog provides high levels of extensibility and allows for a greater degree of flexibility in defining contractual agreements. We have evaluated the approach with respect to complexity and expressiveness by means of experimental tests. In summary, com-

Paschke, A. and Bichler, M.: Knowledge Representation Concepts for Automated SLA Management, Int. Journal of Decision Support Systems (DSS), submitted 19[th] March 2006.

pared to traditional systems, The RBSLM tool[5] serves as a proof-of-concept implementation and illustrates that this particular combination of techniques allows for an efficient and scalable implementation of service level management tools.

---

[5] RBSLA/ ContractLog project site: *http://ibis.in.tum.de/research/projects/rbsla/index.htm*. The RBSLA distribution with the ContractLog framework and the RBSLA language can be downloaded from Sourceforge at: *https://sourceforge.net/projects/rbsla*

Paschke, A. and Bichler, M.: Knowledge Representation Concepts for Automated SLA Management, Int. Journal of Decision Support Systems (DSS), submitted 19th March 2006.

Paschke, A. and Bichler, M.: Knowledge Representation Concepts for Automated SLA Management, Int. Journal of Decision Support Systems (DSS), submitted 19[th] March 2006.